\documentclass[preprint,showpacs,superscriptaddress,12pt]{revtex4-1} %% 

\usepackage{amsmath,amssymb,amsfonts}
\usepackage[english]{babel}
\usepackage{graphicx}

\begin{document}

\title{Optical simulation of Majorana physics}

\author{B. M. Rodr\'{\i}guez-Lara}
\affiliation{Instituto Nacional de Astrof\'{i}sica, \'{O}ptica y Electr\'{o}nica \\ Calle Luis Enrique Erro No. 1, Sta. Ma. Tonantzintla, Pue. CP 72840, M\'{e}xico}
\email{bmlara@inaoep.mx}

\author{H. M. Moya-Cessa}
\affiliation{Instituto Nacional de Astrof\'{i}sica, \'{O}ptica y Electr\'{o}nica \\ Calle Luis Enrique Erro No. 1, Sta. Ma. Tonantzintla, Pue. CP 72840, M\'{e}xico}

\begin{abstract}
We show a procedure to classically simulate the Majorana equation in 1+1 dimensions via two one-dimensional photonic crystals.
We use a decomposition of the Majorana equation into two Dirac equations and propose a novel approach that uses a bi-chromatic refractive index distribution and nearest neighbor couplings of the type found in Glauber-Fock lattices.
This allows us to escape the restriction of staying near the Brillouin zone imposed by the classical simulation of Dirac dynamics with bi-chromatic lattices.
Furthermore, it is possible to simulate the evolution of Gaussian wavepackets under the Majorana/Dirac equation with light impinging only into the first waveguide of our bi-chromatic-Glauber-Fock lattice.
\end{abstract}

\pacs{42.79.Gn, 42.81.Qb, 42.82.Et, 03.65.Pm }
%\ocis{ (000.1600) Classical and quantum physics; (230.4555) Coupled Resonators;  (230.5298) Photonic Crystals; (230.7370) Waveguides;  (350.5500) Propagation;}

\maketitle

\section{Introduction} \label{sec:S1}

Arrays of coupled photonic waveguides have become excellent classical simulators of quantum and relativistic physics; cf. \cite{Longhi2009p243,Longhi2011p453} and references therein.
In particular, they have allowed one of the first visualizations of Dirac physics phenomena like Zitterbewegung and Klein tunneling in tabletop experiments \cite{Dreisow2010p143902,Dreisow2012p10008}.
The Majorana equation is a Lorentz covariant generalization of the Dirac equation \cite{Majorana1932p335,Majorana1937p171}.
It is the first relativistically invariant theory and the first application of the infinite dimensional representations of the Lorentz group \cite{Fradkin1966p314}.
If supplemented by the Majorana condition of charge invariance, i.e. Majorana fermions, it is equivalent to the Dirac equation.
In the standard model the neutrino is the only candidate to be described by the Majorana equation and, for this reason, most interest in its unrestricted version has been purely theoretical.
In modern gauge theories, chiral spinors answer to two-component Majorana equations \cite{Aste2010p1776,Pal2011p485}.
This has rekindled the interest in Majorana dynamics and, recently, it has been proposed both a procedure to implement nonphysical operations related to Majorana physics \cite{Casanova2011p021018} and a scheme to simulate the Majorana equation \cite{Noh2013p040102} with trapped ions setups.
These were the first proposals to bring Majorana physics into the quantum optics laboratory up to our knowledge.

Here we go beyond the quantum optics realizations and show that it is posible to simulate Majorana physics by propagation of classical light in one-dimensional photonic crystals.
For this reason we will present the quantum simulation of the Majorana equation in 1+1 dimension in the following section to provide the basic set of transformations that will move us to/from its classical simulation.
Then, we will propose a photonic analogue in the form of two novel one-dimensional photonic crystals that simulate the Majorana dynamics via the propagation of two initial light distributions for each one.
In our proposal we use a combination of experimentally demonstrated bichromatic \cite{Dreisow2010p143902} and Glauber-Fock \cite{Keil2012p3801} lattices.
This bichromatic-Glauber-Fock lattice allows us to evade the restriction of staying near the Brillouin zone that appears when using bichromatic lattices to classically simulate Dirac physics.
Our proposal imposes no restriction and manages to classically simulate the evolution of Gaussian wavefunctions under Dirac dynamics by a single beam of light impinging the first waveguide of the array.
Finally, we will show that by exploiting the quantum-classical analogue it is straightforward to construct the impulse functions of our lattices.

%%%%%%%%%%%%%%%%%%%%%%%%%%%%%%%%%%%%%%%%%%%%%%%%%%%%%%%%%%%%%%%%%%%%%%%%%%%%%%%%%%%%%%%%%%%%%%%%
\section{Quantum simulation of Majorana physics} \label{sec:S2}
%%%%%%%%%%%%%%%%%%%%%%%%%%%%%%%%%%%%%%%%%%%%%%%%%%%%%%%%%%%%%%%%%%%%%%%%%%%%%%%%%%%%%%%%%%%%%%%%
One of us has been part of an effort to implement a quantum simulation of the Majorana equation in a trapped ion setup \cite{Noh2013p040102}.
Here, we give a brief summary of the topic in order to provide the basis to construct a classical analogue.
We start from the Majorana equation \cite{Majorana1932p335,Majorana1937p171},
\begin{eqnarray} 
i \gamma^\mu \partial_\mu \psi = m_{M} \psi_c,
\end{eqnarray}
where the symbol $\psi_c$ stands for charge conjugation of the spinor $\psi$, $\psi_{c} \equiv \gamma^2 \psi$, and $\gamma_{\mu}$ are the Dirac matrices, the symbol $\partial_{\mu}$ is shorthand notation for partial derivation with respect to $\mu$, $m_{M}$ is the Majorana mass and we have set $\hbar=c=1$.
In 1+1 dimensions it is possible to rewrite it as
\begin{eqnarray}
i \partial_{t} \psi = \left( \hat{\sigma}_{x} \hat{p}_{q} - i m \hat{\sigma}_{y} \right) \psi,
\end{eqnarray}
where the symbols $\hat{\sigma}_{i}$ with $i=x,y,z$ are the Pauli matrices, $\hat{p}_{q}$ is the dimensionless momentum, $m$ stands for the modified Majorana mass in units of momentum, and the Majorana field is given as a two-dimensional complex vector, $\psi = (\psi^{(1)} , \psi^{(2)})$.
Then, we can combine the field and its charge conjugate,
\begin{eqnarray}
i  \partial_{t} \left( \psi + \psi^{\ast} \right) =  \hat{\sigma}_{x} \hat{p}_{q} \left( \psi + \psi^{\ast} \right) + i m  \hat{\sigma}_{y} \left( \psi - \psi^{\ast} \right),\\
i  \partial_{t} \left( \psi - \psi^{\ast} \right) =  \hat{\sigma}_{x} p_{q} \left( \psi - \psi^{\ast} \right) - i m  \hat{\sigma}_{y} \left( \psi + \psi^{\ast} \right), 
\end{eqnarray}
to create and extended real Hilbert space where we can write the Majorana equation as a Schrodinger equation,
\begin{eqnarray}
i \partial_{t} \Psi = \left[ \left( 1_{2} \otimes \hat{\sigma}_{x} \right) \hat{p}_{q} - m  ( \hat{\sigma}_{x} \otimes \hat{\sigma}_{y} ) \right] \Psi
\end{eqnarray}
 with the real four-element field given by $\Psi = ( \mathrm{Re}(\psi^{(1)}),\mathrm{Re}(\psi^{(2)}),\mathrm{Im}(\psi^{(1)}),\mathrm{Im}(\psi^{(2)})  )$ which is related to the Majorana field via the transformation $\psi = \hat{M} \Psi$ with $\hat{M} = ( 1_{2} , i 1_{2})$ where the symbol $1_{2}$ stands for the unit matrix in dimension two.
Note that the unitary operation $\hat{U} = e^{-i \pi \hat{\sigma}_{y} / 4 } \otimes e^{-i \pi \hat{\sigma}_{x} / 4 }$ providing the basis $\Psi = \hat{U} \Phi$, such that we can write the original Majorana field as $\psi = \hat{M} \hat{U} \Phi$, yields the following Schr\"odinger equation related to the Majorana equation:
\begin{eqnarray}
i \partial_{t} \Phi &=& \left[  \left( 1_{2} \otimes \hat{\sigma}_{x} \right) p_{q} + m ( \hat{\sigma}_{z} \otimes \hat{\sigma}_{z} )  \right] \Phi.
\end{eqnarray}
At this point, we can propose a form for our four-vector and realize that this Schr\"odinger representation for the Majorana equation leads to two uncoupled Dirac-like equations with positive and negative mass,
\begin{eqnarray}
i \partial_{t} \phi_{\pm} = \left[ -i \hat{\sigma}_{x} \hat{p}_{q} \pm m \hat{\sigma}_{z} \right] \phi_{\pm}. \label{eq:MajoranaDirac}
\end{eqnarray}
where $\phi_{+} = ( \Phi^{(1)},\Phi^{(2)} )$ and $\phi_{-} = ( \Phi^{(3)},\Phi^{(4)} )$ such that $\Phi = ( \phi_{+}, \phi_{-} )$.

%%%%%%%%%%%%%%%%%%%%%%%%%%%%%%%%%%%%%%%%%%%%%%%%%%%%%%%%%%%%%%%%%%%%%%%%%%%%%%%%%%%%%%%%%%%%%%%%
\section{Photonic lattice analogue} \label{sec:S3}
%%%%%%%%%%%%%%%%%%%%%%%%%%%%%%%%%%%%%%%%%%%%%%%%%%%%%%%%%%%%%%%%%%%%%%%%%%%%%%%%%%%%%%%%%%%%%%%%

Starting from the separation of the Majorana equation into two Dirac equations in \eqref{eq:MajoranaDirac}, we could follow the idea in \cite{Longhi2010p235} and compare the dispersion relation for a bi-chromatic lattice with the energy-momentum dispersion relation of the Dirac equation and conclude that the Majorana Hamiltonian can be simulated classically by two bi-chromatic refractive index lattices described by the differential equation set
\begin{eqnarray}
-i \partial_{z} \mathcal{E}_{j} = \left[ n \pm (-1)^{j} m \right]\mathcal{E}_{j} + \mathcal{E}_{j+1} + \mathcal{E}_{j-1},
\end{eqnarray}
as long as the simulation stays near the boundary of the Brillouin zone.
Also, we have introduced a bias refractive index $n$ that only introduces an overall phase factor.

We don't want such a strong restriction between the momentum of the simulated wave-packet and the characteristics of the bi-chromatic lattice. 
In order to get rid of this restraint, we  map the adimensional linear momentum operator to a combination of bosonic creation (annihilation), $\hat{a}^{\dagger}$ ($\hat{a}$),  operators, 
\begin{eqnarray}
\hat{p}_{q} = \frac{i}{\sqrt{2}} \left( \hat{a}^{\dagger} - \hat{a} \right).
\end{eqnarray} 
Then, we can rewrite the two uncoupled Dirac equations that quantum simulate the Majorana equation as a Schr\"odinger equation with effective Hamiltonians:
\begin{eqnarray}
H_{\pm} = \frac{1}{\sqrt{2}}\left( \hat{a}^{\dagger} + \hat{a} \right) \hat{\sigma}_{x} \pm m \hat{\sigma}_{z}, \label{eq:EffHam}
\end{eqnarray}
after a $\pi/ 2$ rotation around the $\hat{a}^{\dagger}\hat{a}$-axis, $\hat{R} = e^{i \pi  \hat{a}^{\dagger} \hat{a} /2}$ providing a new basis $\varphi$ such that $\phi = \hat{R} \varphi$.
It is straightforward to notice that \eqref{eq:EffHam} is equivalent to the Rabi Hamiltonian \cite{Braak2011p100401} with null field frequency. 
Thus, we can setup a photonic lattice analogue following previous work on the classical simulation of the Rabi Hamiltonian \cite{Crespi2012p163601,RodriguezLara2013p12888}.
In summary, if we split the corresponding Hilbert space into two parity subspaces given by
\begin{eqnarray}
\left\{\vert +, j \rangle \right\}  &=& \left\{ \vert g, 0 \rangle, \vert e, 1 \rangle, \vert g, 2 \rangle, \ldots \right\}, \\
\left\{\vert -, j \rangle \right\}  &=& \left\{ \vert e, 0 \rangle, \vert g, 1 \rangle, \vert e, 2 \rangle, \ldots \right\}, 
\end{eqnarray}
we can write the evolution of any given initial state $\vert \varphi_{\pm, \pm} \rangle = \sum_{j} \mathcal{E}_{\pm, \pm, j} \vert \pm, j \rangle$, where the first subindex is related to the positive/negative mass Hamiltonian and the second to the positive/negative parity, as the vector differential set
\begin{eqnarray}
i \partial_{t} E_{\pm,\pm} = H_{\pm,\pm} E_{\pm,\pm} , \label{eq:MatrixDifSet}
\end{eqnarray}
where the vector of amplitudes is given by $E_{a,b} = \left( \mathcal{E}_{a,b,0}, \mathcal{E}_{a,b,1}, \ldots \right)$ and the elements of the four matrices describing the dynamics reduce to two matrices as
\begin{eqnarray}
\left( H_{+,+} \right)_{j,k} &=& \left( H_{-,-} \right)_{j,k} \\
&=&  -m(-1)^{j} \delta_{j,k} + \sqrt{\frac{k}{2}} \delta_{j+1,k} + \sqrt{\frac{j}{2}} \delta_{j-1,k}, \\
\left( H_{+,-} \right)_{j,k} &=& \left( H_{-,+} \right)_{j,k}  \\
&=&  m(-1)^{j} \delta_{j,k} + \sqrt{\frac{k}{2}} \delta_{j+1,k} + \sqrt{\frac{j}{2}} \delta_{j-1,k}.
\end{eqnarray}
By making the change $t \rightarrow z$ we obtain, up to a constant phase factor, a differential set describing two photonic lattices where the individual refractive indices are bi-chromatic \cite{RodriguezLara2013p038116} and the nearest neighbour couplings go as those in Glauber-Fock photonic lattices \cite{RodriguezLara2011p053845,PerezLeija2012p013848}.
It is straightforward to realize that the magnitude of the effective mass $m$ will provide three dynamics regimes: $ m \gg 1/\sqrt{2}$,  $m \sim 1/\sqrt{2}$ and  $m \ll 1/\sqrt{2}$. 

\begin{figure}
\includegraphics[scale=1]{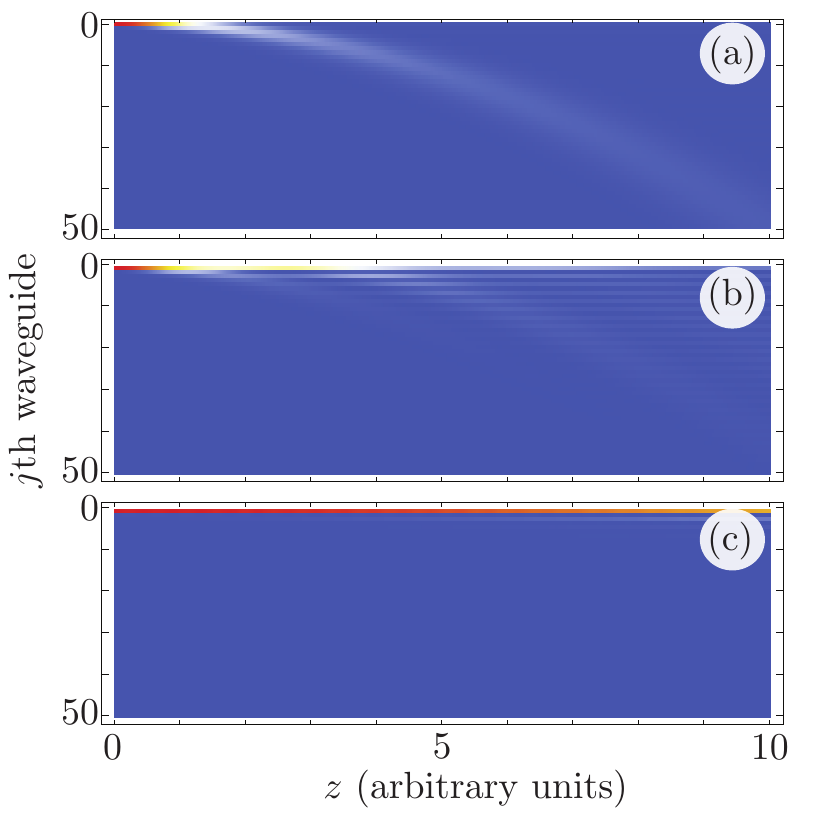}
\caption {(Color online) The time evolution of the modulus squared of an initial Gaussian wavepacket, $\Psi_{0}(q) = e^{-q^2 / 2} / \pi^{1/4}$, simulated by the propagation of light impinging the first waveguide of an array of photonic waveguides described by the matrix $H_{+,+}$ or $H_{-,-}$ with effective mass $m= 1/\sqrt{2} \times$ (a) $0.1$, (b) $1$ and (c) $10$.}  \label{fig:Fig1}
\end{figure}

\begin{figure}
\includegraphics[scale=1]{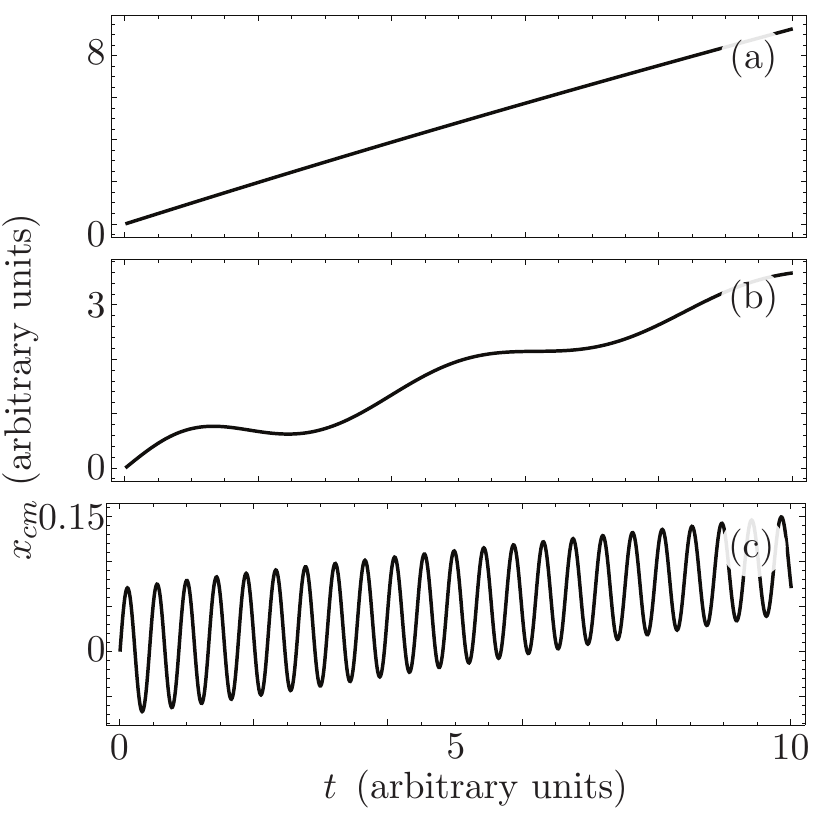}
\caption {(Color online) The evolution of the center of mass for an initial Gaussian wavepacket, $\Psi_{0}(q) = e^{-q^2 / 2} / \pi^{1/4}$, under Dirac dynamics recovered from the propagation of an initial beam impinging the first waveguide in an array of photonic waveguides described by the matrix $H_{+,+}$ or $H_{-,-}$ related to effective mass parameters $m= 1/\sqrt{2} \times$ (a) $0.1$, (b) $1$ and (c) $10$ described in Fig \ref{fig:Fig1}.}  \label{fig:Fig2}
\end{figure}

Before advancing further, we want to discuss the meaning of light impinging the $j$th waveguide in one of our photonic lattices which, after dropping two of the subsindices refering to the parity and the mass sign,  is equivalent to writing 
\begin{eqnarray}
\vert \psi_{j}(0) \rangle &=& \vert j \rangle, \\
&=& \int dq ~ \vert q \rangle ~ \Psi_{j}(q) ,
\end{eqnarray}
with 
\begin{eqnarray}
\Psi_{j}(q) &\equiv& \langle q \vert j \rangle, \\
&=& \frac{1}{\sqrt{2^{j} j!}} \left( \frac{1}{\pi} \right)^{1/4} e^{-q^2/2} H_{j}(q) ,  \label{eq:ONBase}
\end{eqnarray}
where $H_{n}(x)$ is the $n$th Hermite polynomial.
In other words, a beam of light impinging just the first waveguide of the array is equivalent to an initial Gaussian wavefunction in dimensionless canonical space, $\Psi_{0}(q) = e^{-q^{2}/2} / \pi^{1/4}$. 
Light impinging just the $j$th waveguide will simulate an initial wavefunction in canonical space with a $j$th Hermite-Gaussian distribution.
Any given initial wavefunction in dimensionless canonical space, $\psi(q)$, can be constructed from an adequate superposition of classical fields impinging the photonic crystal, 
\begin{eqnarray}
\psi(q) = \sum_{j} \mathcal{E}_{j}(0) \Psi_{j}
\end{eqnarray}
where the initial field amplitudes at each waveguide, $\mathcal{E}_{j}(0)$, correspond to the decomposition of the initial wavefunction in the orthonormal basis provided by (\ref{eq:ONBase}),
\begin{eqnarray}
\mathcal{E}_{j}(0) = \int dq ~\psi(q) \Psi_{j}^{\ast}(q).
\end{eqnarray}
Note that this has to be done for each set of initial conditions to be propagated in each photonic crystal.
Figure \ref{fig:Fig1} shows the numerical propagation of a Gaussian wavepacket in canonical space, $\Psi_{0}$, simulated by a beam of light impinging the first waveguide, when the effective mass belongs to different regimes: Fig. \ref{fig:Fig1}(a) $m= 0.1/\sqrt{2}$, Fig. \ref{fig:Fig1}(b) $m= 1/\sqrt{2}$ and  Fig. \ref{fig:Fig1}(c) $m= 10/\sqrt{2}$. 
Note that for an initial Gaussian packet as the effective mass becomes negligible, $m \rightarrow 0$, the amplitudes distribution will become closer to a Poisson distribution with mean and variance given by $t^2/2$, $\psi(q,t) = \sum_{j} e^{-t^2/4} (i t)^{j} / \sqrt{j!}$ \cite{PerezLeija2010p2409,RodriguezLara2011p053845}, while in the complete opposite case, $m \rightarrow \infty$, the amplitude distribution will become propagation invariant.
%
%%%%%%%%%%%%%%%%%%%%%%%%%%%%%%%%%%%%%%%%%%%%%%%%%%%%%%%%%%%%%%%%%%%%%%%%%%%%%%%%%%%%%%%%%%%%%%%%
%%%%%%%%%%%%%%%%%%%%%%%%%%%%%%%%%%%%%%%%%%%%%%%%%%%%%%%%%%%%%%%%%%%%%%%%%%%%%%%%%%%%%%%%%%%%%%%%
With our classical simulation we have direct access to the center of mass of the intensity distribution for variable effective mass parameters.
Please, be aware that in our simulation the center of mass for the intensity does not coincide with the center of mass of the wavefunction in canonical space,
\begin{eqnarray}
q_{cm} &=& \sum_{k} \sqrt{\frac{k+1}{2}} ~\left[ \mathcal{E}_{k}^{\ast}(t) \mathcal{E}_{k+1}(t) + \mathcal{E}_{k+1}^{\ast}(t) \mathcal{E}_{k}(t) \right],
\end{eqnarray}
and, furthermore, in order to study the center of mass under the original dynamics we have to account for all the rotations to transform back to the original frame, $x_{cm}$; e.g. for the Dirac dynamics part of our classical simulation:
\begin{eqnarray}
x_{cm} = \sum_{k} i \sqrt{\frac{k+1}{2}} ~\left[ \mathcal{E}_{k}^{\ast}(t) \mathcal{E}_{k+1}(t) - \mathcal{E}_{k+1}^{\ast}(t) \mathcal{E}_{k}(t) \right].
\end{eqnarray} 
Figure \ref{fig:Fig2} shows the evolution of the center of mass of an initial Gaussian wavepacket under Dirac dynamics for different effective masses, Fig. \ref{fig:Fig1}(a) $m= 0.1/\sqrt{2}$, Fig. \ref{fig:Fig1}(b) $m= 1/\sqrt{2}$ and  Fig. \ref{fig:Fig1}(c) $m= 10/\sqrt{2}$, recovered from the classical simulation equivalent to the propagation of a beam impinging the first waveguide of the photonic crystal.

%%%%%%%%%%%%%%%%%%%%%%%%%%%%%%%%%%%%%%%%%%%%%%%%%%%%%%%%%%%%%%%%%%%%%%%%%%%%%%%%%%%%%%%%%%%%%%%%
%%%%%%%%%%%%%%%%%%%%%%%%%%%%%%%%%%%%%%%%%%%%%%%%%%%%%%%%%%%%%%%%%%%%%%%%%%%%%%%%%%%%%%%%%%%%%%%%

We are also interested in providing an impulse function for our photonic crystal.
For this reason we start from the matrix diferential set \eqref{eq:MatrixDifSet} where the matrix $H_{a,b}$ is constant and allows us to write the time propagator as 
\begin{eqnarray}
U(t) &=& e^{-i H_{a,b} t}, \\
&=& \cos \Omega(\hat{q}) t - \frac{i}{\Omega(\hat{q})} \sin \Omega(\hat{q}) t ~ H_{a,b},
\end{eqnarray}
with 
\begin{eqnarray}
\Omega(\hat{q}) &=& \sqrt{ m^{2} + \frac{1}{2}\left( \hat{a} + \hat{a}^{\dagger} \right)^2}, \\
 &=& \sqrt{ m^{2} + \hat{q}^2}.
\end{eqnarray}
where $\hat{q}$ is the dimensionless canonical position operator.
This expression is related to the dispersion relation of the photonic lattice and the energy states of the Dirac equation.
These results and the action of the matrix $H_{a,b}$ over the components of the parity basis allow us to calculate the impulse function (field amplitude) at the $k$th waveguide for an input in the $j$th waveguide of the corresponding photonic lattice; e.g. for the waveguide array described by the matrices $H_{+,+}$ and $H_{-,-}$ we can write the impulse function as:
\begin{eqnarray}
I^{(+,+)}_{j,k} &=&  I^{(-,-)}_{j,k}, \\
&=& \int dq ~ \left\{ \cos \Omega(q) t - \frac{i \left[ - m (-1)^{j} + q \right]}{\Omega(q)} \sin \Omega(q) t  \right\} \Psi^{\ast}_{k}(q) \Psi_{j}(q). \label{eq:ImpulseFunction}
\end{eqnarray}
Comparing this expression with numerical propagation in the photonic lattice shows good agreement between the results.
It's straightforward to use the impulse function to calculate the propagation of any initial light field simulating any given initial Majorana field.

%%%%%%%%%%%%%%%%%%%%%%%%%%%%%%%%%%%%%%%%%%%%%%%%%%%%%%%%%%%%%%%%%%%%%%%%%%%%%%%%%%%%%%%%%%%%%%%%
%\begin{figure}
%\includegraphics[scale=1]{Fig3.pdf}
%\caption {(Color online) The light intensity at the first waveguide when the initial %condition is a beam impinging the first waveguide of an array of photonic waveguides %described by the matrix $H_{+,+}$ or $H_{-,-}$ with effective mass $m= 1/\sqrt{2} %\times$ (a) $0.1$, (b) $1$ and (c) $10$. The solid (black) line comes from the numerical %propagation and the dashed (red) line from the squared modulus of the impulse function %in \eqref{eq:ImpulseFunction}. }  \label{fig:Fig3}
%\end{figure}
%%%%%%%%%%%%%%%%%%%%%%%%%%%%%%%%%%%%%%%%%%%%%%%%%%%%%%%%%%%%%%%%%%%%%%%%%%%%%%%%%%%%%%%%%%%%%%%%

%%%%%%%%%%%%%%%%%%%%%%%%%%%%%%%%%%%%%%%%%%%%%%%%%%%%%%%%%%%%%%%%%%%%%%%%%%%%%%%%%%%%%%%%%%%%%%%%
\section{Conclusions} \label{sec:S6}
%%%%%%%%%%%%%%%%%%%%%%%%%%%%%%%%%%%%%%%%%%%%%%%%%%%%%%%%%%%%%%%%%%%%%%%%%%%%%%%%%%%%%%%%%%%%%%%%
We have shown a scheme to classical simulate Majorana physics in 1+1 dimension with two one-dimensional photonic crystals.
Our approach is based in the quantum simulation of the Majorana equation through two Dirac equations with positive/negative mass. 
We have proposed a novel way to classically simulate the Dirac equation combining two one-dimensional photonic crystals that have already been produced experimentally; one is the bi-chromatic photonic lattice and the other is the Glauber-Fock photonic lattice.
Each Dirac equation has an optical analogue in a set of two bichromatic-Glauber-Fock lattices; i.e. a waveguide array where the refractive index of individual waveguides alternate and the coupling goes as the square root of the waveguide number.
We have also demonstrated that instead of using four photonic lattices it is enough to use two waveguide arrays with the propagation of two adequate initial conditions in each one to simulate the Majorana equation.
Finally, by using the equivalence between the quantum and classical simulation we were able to give the impulse function of the photonic lattices.
These impulse functions allow us to calculate the propagation of any initial light distribution simulating any given Majorana field.

%%%%%%%%%%%%%%%%%%%%%%%%%%%%%%%%%%%%%%%%%%%%%%%%%%%%%%%%%%%%%%%%%%%%%%%%%%%%%%%%%%%%%%%%%%%%%%%%
\begin{acknowledgments}
BMRL is grateful to Changsuk Noh for valuable comments.
\end{acknowledgments}
%%%%%%%%%%%%%%%%%%%%%%%%%%%%%%%%%%%%%%%%%%%%%%%%%%%%%%%%%%%%%%%%%%%%%%%%%%%%%%%%%%%%%%%%%%%%%%%%

%%%%%%%%%%%%%%%%%%%%%%%%%%%%%%%%%%%%%%%%%%%%%%%%%%%%%%%%%%%%%%%%%%%%%%%%%%%%%%%%%%%%%%%%%%%%%%%%
%\bibliography{D:/ExternalHD/Bibliography/references}

\end{document}